\documentclass{svjour3}
\usepackage{geometry}
\renewcommand{\qed}{\hfill\rule{1ex}{1ex}}

\usepackage[normalem]{ulem}
\usepackage{tikz}

\usepackage{amssymb,amsmath,amsfonts,amssymb,amsthm}
\usepackage{bbm}
\usepackage{float}
\usepackage{graphicx}
\usepackage{paralist}
\usepackage{floatflt}
\usepackage{alltt}
\usepackage{algorithm,algorithmic}
\usepackage{hyperref}
\usepackage{color,xcolor}
\definecolor{string}{rgb}{0.7,0.0,0.0}
\definecolor{comment}{rgb}{0.13,0.54,0.13}
\definecolor{keyword}{rgb}{0.0,0.0,1.0}

\DeclareMathOperator{\proj}{proj}
\DeclareMathOperator{\conv}{conv}

\begin{document}

\large
\renewcommand{\arraystretch}{0.5}

\title{Warehouse Problem with Time-varying Bounds, Fixed Costs and Complementary Constraints\thanks{
		This work  was partially supported by ONR  Grant N00014-211-2575.. }
}
\author{Ishan Bansal, Oktay G\"unl\"uk}
\institute{\\   \and    
	Oktay G{\"u}nl{\"u}k,    Georgia Tech,   \email{ogunluk3@gatech.edu} \\\and
	Ishan Bansal,   Amazon Bellevue   \email{ishanbap@amazon.com} {~~}(This work was completed prior to being affiliated with Amazon and does not relate to the current role at Amazon)
}

\date{\today}


\newcommand{\matlab}{\textsc{Matlab\;}}
\newcommand{\grads}{[\text{565 only}]}
\newcommand{\maple}{\textit{Maple\;}}
\newcommand{\erf}{\text{erf}}
\newcommand{\sign}{\text{sign}}
\newcommand{\lnorm}{\left|}
\newcommand{\rnorm}{\right|}
\newcommand{\R}{\mathbb{R}}
\newcommand{\x}{\ensuremath{x}}
\newcommand{\y}{\ensuremath{y}}
\newcommand{\s}{\ensuremath{s}}
\newcommand{\w}{\ensuremath{w}}
\newcommand{\z}{\ensuremath{z}}
\newcommand{\f}{\ensuremath{f}}
\newcommand{\snought}{\ensuremath{s_0}}
\newcommand{\p}{\ensuremath{p}}
\newcommand{\cost}{\ensuremath{c}}
\newcommand{\g}{\ensuremath{g}}
\newcommand{\h}{\ensuremath{h}}
\newcommand{\Us}{\ensuremath{U^s}}
\newcommand{\Uy}{\ensuremath{U^y}}
\newcommand{\Ux}{\ensuremath{U^x}}
\newcommand{\xt}{\ensuremath{x_t}}
\newcommand{\yt}{\ensuremath{y_t}}
\newcommand{\st}{\ensuremath{s_t}}
\newcommand{\wt}{\ensuremath{w_t}}
\newcommand{\zt}{\ensuremath{z_t}}
\newcommand{\pt}{\ensuremath{p_t}}
\newcommand{\ct}{\ensuremath{c_t}}
\newcommand{\rt}{\ensuremath{r_t}}
\newcommand{\gt}{\ensuremath{g_t}}
\newcommand{\htt}{\ensuremath{h_t}}
\newcommand{\Ust}{\ensuremath{U^s_t}}
\newcommand{\Uyt}{\ensuremath{U^y_t}}\newcommand{\Uxt}{\ensuremath{U^x_t}}
\newcommand{\Uyi}{\ensuremath{U^y_i}}\newcommand{\Uxi}{\ensuremath{U^x_i}}
\newcommand{\ft}{\ensuremath{f_t}}
\newcommand{\tbound}{\ensuremath{T}}
\newcommand{\tset}{\mathcal{T}}
\newcommand{\Pone}{\ensuremath{P_0}}
\newcommand{\Ptwo}{\ensuremath{P_1}}
\newcommand{\Pthree}{\ensuremath{P_2}}
\newcommand{\Qtwo}{\ensuremath{Q_1}}
\newcommand{\Qthree}{\ensuremath{Q_2}}
\newcommand{\tsell}{\ensuremath{t^{\text{sell}}}}\newcommand{\tbuy}{\ensuremath{t^{\text{buy}}}}
\renewcommand{\tsell}{\ensuremath{t^{-}}}\renewcommand{\tbuy}{\ensuremath{t^{+}}}
\newcommand{\Tsell}{\ensuremath{T^{-}}}\newcommand{\Tbuy}{\ensuremath{T^{+}}}
\newcommand{\stsell}{\ensuremath{\s_{\tsell}}}
\newcommand{\stbuy}{\ensuremath{\s_{\tbuy}}}
\newcommand{\Tspl}{\mathcal{T}_{spl}}
\newcommand{\potstockbound}{S}
\newcommand{\skipthis}[1]{}
\newcommand{\oo}[1]{{\textcolor{purple}{#1}}}
\newcommand{\ooc}[1]{\textcolor{purple}{[OG comment: \oo{#1}]} }
\newcommand{\stoo}[1]{\textcolor{blue}{[\sout{#1}]} }
\newcommand{\set}{,\quad}
\newcommand{\ignore}[1]{#1}
\maketitle

\begin{abstract}
	The warehouse problem is a classical problem arising in revenue management in a commodity market. The objective is to decide on optimal quantities of the commodity to purchase, store and sell when the market purchase and sale prices change over time. In this paper, we study deterministic variants of the warehouse problem with time-varying bounds on purchase, storage, and sales quantities, fixed costs for purchasing and selling, and complementarity constraints that prohibit purchases and sales during the same time period.	
	We characterize the extreme points of the linear warehouse polytope with time-varying bounds and exploit this characterization to construct compact network-flow formulations for the fixed-cost and complementarity-constrained models.
	These formulations yield pseudo-polynomial-time algorithms and extended linear formulations. Although the general problems are NP-hard, we obtain polynomial-time algorithms for important special cases, including time-independent bounds and fixed-dimensional lattice-structured time-varying bounds. 
	Finally, we show that our framework extends naturally to models with lower bounds, storage costs, and separable convex pay-off functions.
\end{abstract}

%

\section{Introduction}\label{sec:intro}

In this paper, we study the warehouse problem which involves deciding on purchase and sale quantities of a single product over a finite (discrete) time horizon. 
This problem was originally introduced and studied by Cahn~\cite{Cahn} where the author considers a given (fixed) warehouse capacity for storage and optimizes the profit when purchase and  sale prices change over time. The original motivation for this problem comes from commodity markets involving raw materials or agricultural products (see, e.g.,~\cite{devalkar,secomandi,geman}) where an agent maximizes total profit over a finite time horizon exploiting the changes in prices.
More recent applications of this problem involve energy markets where agents can produce or buy electricity, and store it using various devices and technologies and sell it when profitable. There is growing literature on the operation of consumer-owned electric storage systems (see, e.g.,~\cite{xuconference,xujournal}) and also the so-called energy arbitrage problem (see, e.g.,~\cite{bolun2020,bolun2021}). Most of these applications model the problem as a stochastic variant of the original warehouse problem and solution methods are often based on dynamic programming.

The original warehouse problem has been extended in two main directions by incorporating multiple items (that share a common storage facility) and by considering the uncertainty around (future) prices and therefore turning this deterministic problem into a stochastic (or robust) optimization problem.
It was shown by Charnes et al.~\cite{charnes} that the stochastic version of the problem as stated above can be reduced to the deterministic version where the selling and purchasing costs of the commodity over time are given by their expected values.

In this paper, we focus on the deterministic version of the problem. As in
classical warehouse models, some stochastic variants can be handled by replacing
future prices or payoffs by their expected values when these expectations are
known or can be computed efficiently.
More precisely, we consider three variants of the basic warehouse problem that involve  time-varying bounds  on purchase, storage, and sales quantities. In addition, we consider fixed costs for purchases and sales as well as complementarity constraints that prohibit purchases and sales during the same time period.
Many existing algorithms for warehouse-type problems are based on dynamic programming and rely on concavity properties of the dynamic programming value functions;  see \cite{bellman,dreyfus,secomandi,devalkar,xujournal} and related energy-arbitrage	models~\cite{bolun2020,bolun2021}. 	
	Fixed costs and complementarity	constraints destroy this property, which is why they are not handled by the
	standard dynamic-programming framework.
	Our approach is instead polyhedral: we characterize the extreme points of the linear warehouse polytope with time-varying bounds and use this characterization to build network-flow formulations for the fixed-cost and complementarity-constrained variants.
We next formally describe three problems that we study in this paper and give integer programming formulations for them.

\subsection{Problem Definition}
In all three problems we use the following notation. Let $\tset=\{1,\ldots,\tbound\}$ denote the planning periods where $\tbound$ is the  number of periods in the  horizon. During each time period $t\in\tset$, let $\ct$ be the cost per unit for purchases, $\pt$ be the cost per unit for sales, and $\rt$ be the unit cost of storage. Note that the costs $\ct$, $\pt$, and $\rt$ need not be positive. Let $\Ust$ be the bound on storage quantities or stock at the end of time period $t\in\tset$, and similarly, let $\Uxt$ and $\Uyt$ denote the bounds on purchase and sales quantities, respectively. Let $\snought$ be the initial stock in the warehouse at the beginning of the planning horizon. 
Following a classical and practically relevant assumption in the warehouse problem, we assume that within each time period all sales occur before any purchases are made. Consequently, the quantity sold in a period cannot exceed the stock available at the beginning of that period.

The first variant of the warehouse problem that we consider extends the original model by allowing time-dependent bounds on purchase, sale  and storage quantities. To formulate this problem as a linear program (LP), we introduce variables $\st$ to model the stock at the end of time period $t$. Similarly, we introduce variables $\xt$ and $\yt$ to model the purchase and sale quantities, respectively, during time period $t$.  We can now model the problem of maximizing profit as the following LP:

\begin{align*}\label{prob:1}
	\max & \;\;\;\;\sum_{t\in\tset} ~(\pt\yt + \ct\xt +\rt\st)\tag{\bf{LP0}}\\
	\text{s.t.}&\;\;\;\; \st = \s_{t-1} - \yt + \xt & \forall\; t\in\tset\\
	&\;\;\;\;\yt \leq \s_{t-1} &\forall\; t\in\tset\\
	&\;\;\;\; 0\leq \st\leq \Ust,\;\;\;\;0\leq \yt\leq \Uyt,\;\;\;\;0\leq \xt\leq \Uxt& \forall\; t\in\tset
\end{align*}

\noindent We refer to the feasible set of the above linear program as $\Pone$. While $\Pone$ can be solved in polynomial time, the variants we consider next are NP-hard. 

\begin{figure}[t]\centering	\begin{tikzpicture}[scale=.04]
		\draw[->] (25*5-15,10) -- (25*5+5,10);\node at (25*6-3,10){$\cdots$};
		\foreach \x in {1,2,3,5.5,6.5}   {\draw  (35*\x,10) circle (5cm); \draw[->] (35*\x,5) -- (35*\x,-10); \draw[<-] (35*\x,15) -- (35*\x,35);\draw[->] (35*\x-30,10) -- (35*\x-5,10);}\draw[->] (35*7.5-30,10) -- (35*7.5-5,10);
		\foreach \x in {1,2,3}   \node at (35*\x+7,-10){$y_\x$};  \node at (35*5.5+19,-10){$y_{n-1}$}; \node at (35*6.5+10,-10){$y_{n}$};
		\foreach \x in {0,1,2,3} \node at (35*\x+15,15){$s_{\x}$};\node at (35*4.5+15,15){$s_{n-2}$};\node at (35*5.5+15,15){$s_{n-1}$};\node at (35*6.5+15,15){$s_{n}$};
		\foreach \x in {1,2,3}   \node at (35*\x+7,35){{$x_\x$}};  \node at (35*5.5+15,35){{$x_{n-1}$}}; \node at (35*6.5+10,35){{$x_{n}$}};
	\end{tikzpicture}\caption{Base model for the warehouse problem}
\end{figure}

In the second variant, we introduce time-dependent fixed costs for purchasing and selling. Let $\gt$ denote the fixed cost of purchasing during time period $t$ and similarly, let $\htt$  denote the fixed cost of selling during time period $t$. Without loss of generality, we assume that  $\htt, \gt\le0$. To model this problem as an integer program (IP), we introduce binary decision variables $\wt$ and $\zt$ with the interpretation that $\wt = 1$ if a purchase takes place during time period $t$, and, $\zt=1$ if a sale takes place. The IP then becomes,

\begin{align*}\label{prob:2}
	\max & \;\;\;\;\sum_{t=1}^\tbound ~(\pt\yt + \ct\xt  +\rt\st +\htt\zt + \gt\wt) \tag{\bf{IP1}}\\
	\text{s.t.}&\;\;\;\; \st = \s_{t-1} - \yt + \xt & \forall\; t\in\tset\\
	&\;\;\;\;\yt \leq \s_{t-1} &\forall\; t\in\tset \\
	&\;\;\;\; 0\leq \st\leq \Ust,\;\;\;\;0\leq \yt\leq \Uyt\zt,\;\;\;\;0\leq \xt\leq \Uxt\wt& \forall\; t\in\tset\\
	&\;\;\;\; \zt,\wt\in\{0,1\} &\forall\; t\in\tset
\end{align*}

\noindent We refer to the feasible set of the above IP as $\Qtwo$ and to its convex hull as $\Ptwo = \conv(\Qtwo)$.

Finally, we consider complementary constraints that prohibit purchasing and selling during the same time period. We model this variant using the following quadratically constrained integer program.

\begin{align*}\label{prob:3}
	\max & \;\;\;\;\sum_{t=1}^\tbound ~(\pt\yt + \ct\xt  +\rt\st + \htt\zt + \gt\wt) \tag{\bf{IP2}}\\
	\text{s.t.}&\;\;\;\; \st = \s_{t-1} - \yt + \xt & \forall\; t\in\tset\\
	&\;\;\;\;\yt\xt = 0 &\forall\; t\in\tset \\
	&\;\;\;\; 0\leq \st\leq \Ust,\;\;\;\;0\leq \yt\leq \Uyt\zt,\;\;\;\;0\leq \xt\leq \Uxt\wt& \forall\; t\in\tset\\
	&\;\;\;\; \zt,\wt\in\{0,1\} &\forall\; t\in\tset
\end{align*}

\noindent We will refer to the feasible set of the above quadratic integer program as $\Qthree$ and to its convex hull as $\Pthree = \conv(\Qthree)$. Note that the complementary constraints above can be replaced by the linear constraint  $\zt+\wt\le1$ when $\htt, \gt\le0$ for all $t\in\tset$.

\subsection{Our Results}\label{subsec:ourresults}

We begin by characterizing the extreme points of the polytope $\Pone$ in Section~\ref{sec:extreme}. This is done by observing that in any extreme point of $\Pone$, it is not possible to purchase or sell at mid-capacity multiple times unless the warehouse reaches full or zero stock. Using this property we identify certain points in $\Pone$ called \textit{critical points}. We show that the set of extreme points of $\Pone$ is equal to the set of critical points of $\Pone$. We also show that critical points of $\Pone$ have useful structural properties that allow us to describe a set of potential stock levels at any time for an extreme point of $\Pone$. This characterization is the key structural result underlying all of the algorithms and extended formulations developed in this paper.

The description of potential stock levels given above is  closely related to the state space used in dynamic programming approaches to the warehouse problem \cite{bellman,dreyfus,secomandi,devalkar}. These approaches are particularly effective when the bounds are fixed or otherwise structured.
In contrast, our characterization is obtained through a polyhedral analysis of the extreme points of $P_0$ and  leads to compact network-flow formulations even when the bounds are time-varying.

In Section \ref{sec:fixedcosts}, we use the results from Section \ref{sec:extreme} and generalize the ideas put forth by Wolsey and Yaman~\cite{wolsey} to obtain an algorithm for Problem \eqref{prob:2} and an extended linear formulation for the polytope $\Ptwo$. 
In Section \ref{sec:comp} we extend these ideas to Problem \eqref{prob:3}.
Problems~\eqref{prob:2} and~\eqref{prob:3} are {NP}-hard problems, and the running times of the algorithms developed in 	Section~\ref{sec:fixedcosts} are exponential in the worst case. 
	
In Section~\ref{sec:special} we identify important special cases for which our algorithms become polynomial-time. In particular, the most common setting in recent warehouse applications assumes that the purchase, storage, and sales bounds are time-independent. We show that if $\Ust=\Us$, $\Uxt=\Ux$, and $\Uyt=\Uy$ for all $t\in\tset$, then Problems~\eqref{prob:2} and~\eqref{prob:3} can be solved in time $O(T^3\gamma^2)$, where $\gamma=\min\left\{T,\frac{\Us}{\max\{\Ux,\Uy\}}\right\}$.
More generally, we show that if the time-varying bounds admit a fixed-dimensional lattice representation, that is, if they can be expressed as $\sum_{i=1}^{k}\alpha_i d_i$, where $d_i>0$ and $\alpha_i\in\{0,1,\ldots,r_i\}$ for some non-negative integers $r_i$, then Problems~\eqref{prob:2} and~\eqref{prob:3} can be solved in time $O((2T)^{2k+1}r_1^2r_2^2\cdots r_k^2)$.
	
Finally, in Section~\ref{sec:further} we observe that our results and methods extend naturally to models with lower bounds on purchases, storage, and sales, as well as fixed costs for storage. 
Lastly, we show that the linear objective functions considered throughout the paper can be replaced by separable convex \textit{pay-off} functions without changing our overall solution approach.

\subsection{Related Work}

The original warehouse problem was introduced by Cahn \cite{Cahn}. This original problem can be seen as a special case of Problem \eqref{prob:1} where there is no storage cost and all the bounds are equal to a fixed constant, i.e., $\Ust = \Uxt = \Uyt = U$ and $\rt=0$ for all $t\in\tset$. Bellman \cite{bellman} presented a dynamic programming algorithm to solve the original problem. Dreyfus \cite{dreyfus} extended this dynamic programming approach and showed that the solution can be determined analytically. Charnes et al. \cite{charnes} considered a variant of the original warehouse problem where the selling and buying costs evolve over time according to a known stochastic process. They showed that results for the deterministic problem can be used to obtain a policy that is optimal in expectation. 

Subsequent work has studied many bounded-storage and commodity-trading variants, including applications to natural gas \cite{secomandi}, agricultural commodities \cite{devalkar}, and energy storage \cite{xuconference,xujournal,bolun2020,bolun2021}. 
These papers typically consider a stochastic variant of Problem \eqref{prob:1} with bounds that are fixed or otherwise structured. The resulting models are often solved using	dynamic-programming approaches that rely on concavity properties of the single-period value functions. 
Complementarity constraints destroy this concavity structure. As a result, papers that include such constraints typically impose assumptions such as $p_t+c_t\le 0$, under which simultaneous purchasing and selling is suboptimal and the complementarity constraints become effectively redundant.

To the best of our knowledge, the only paper that studies fixed costs and complementarity constraints in the warehouse problem is the work of Wolsey and Yaman~\cite{wolsey}. They also consider problems \eqref{prob:2} and \eqref{prob:3}, but assume that all the purchase, storage, and sales bounds are equal to a fixed constant, i.e., $\Ust=\Uxt=\Uyt=U$.
In this setting, the results of Dreyfus~\cite{dreyfus} lead to a description of potential stock levels at any time period in an extreme point of the polytope $\Pone$. Using this, Wolsey and Yaman devised network flow formulations for problems \eqref{prob:2} and \eqref{prob:3} and obtained linear time algorithms to solve them. The methods of Dreyfus~\cite{dreyfus} were also based on dynamic programming and attempts to generalize those methods to arbitrary time-varying bounds do not immediately provide satisfactory descriptions of potential stock levels at extreme points. For example in the setting where $\Ust = \Us,\Uxt = \Ux$ and $\Uyt = \Uy$ for all $t\in\tset$, Devalkar et al.~\cite{devalkar} observed that using dynamic programming, one can show that the set of potential stock levels at the end of any time period in an extreme point of $\Pone$ contains multiples of $\gcd(\Us,\Ux,\Uy)$. This is a pseudo-polynomially sized set. Using our methods, we obtain a set which is strongly polynomial in size. Thus, our novel characterization of extreme points of $\Pone$ along with a generalization of the ideas put forth by Wolsey and Yaman~\cite{wolsey} allows us to obtain our results.

We would also like to point out a closely related problem where the amounts sold during each time period is not a variable but a fixed demand quantity. Thus $\yt = d_t$ for some fixed constant $d_t$. This is a classical problem known as the lot-sizing problem and has been heavily studied. We refer the reader to the book by Pochet and Wolsey~\cite{pochet} for a wide and detailed analysis of the lot-sizing problem. Note that the lot-sizing variant of problem \eqref{prob:2} can be reduced to problem \eqref{prob:2} by setting $\Uyt = d_t$ and setting $p_t = M$ for a large enough constant $M$ so that it is never optimal to sell quantities that are lower than $d_t$. It was shown by Bitran and Yanasse \cite{bitran} that the lot-sizing variant of problem \eqref{prob:2} is {NP}-hard and thus by the above reduction, problem \eqref{prob:2} is itself {NP}-hard. Furthermore, one can reduce problem \eqref{prob:2} to problem \eqref{prob:3} by just splitting each time period into a sell period followed by a buy period and setting the upper bounds of $\xt$ to be zero during a sell period and the upper bound of $\yt$ to be zero during a buy period. Doing this will ensure that $\yt\xt=0$ is vacuously satisfied and so can be added in as a constraint. Hence problem \eqref{prob:3} is also {NP}-hard.

\section{Extreme Points of {$\Pone$}}\label{sec:extreme}

Our approach to solving  problems \eqref{prob:2} and \eqref{prob:3}  consists of three main steps. We first characterize the extreme points of the feasible region. We then show that these extreme points admit a compact network representation. Finally, we exploit this representation to obtain both a pseudo-polynomial-time algorithm and an extended linear formulation.

We start with  characterizing the extreme points of the polytope $\Pone$ by analyzing possible values that the stock level variables $s_t$ can take in an extreme point solution. The main idea here is to show the following property for extreme points of $\Pone$: If the buy or sell quantities are strictly between their bounds at $t\in\tset$ i.e. $0<\xt<\Uxt$ or $0<\yt<\Uyt$, then during subsequent time periods we must keep selling and purchasing at full or zero capacity until some time $t'\geq t$ when the warehouse stock is equal to either $0$ or $\Us_{t'}$. This observation allows us to characterize the extreme points of the polytope $\Pone$. 

To formalize this idea, we split every time period $t\in\tset$ into two consecutive split periods: a \textit{sell period} $\tsell$, during which only sales are allowed, followed by a \textit{buy period} $\tbuy$, during which only purchases are allowed. Let $s_{\tsell}$ and $s_{\tbuy}$ denote the stock level immediately after the sell and buy periods, respectively. Then
$$ s_{\tsell}=s_{t-1}-y_t,\qquad s_{\tbuy}=s_t,$$
and consequently, $0\le s_{\tsell},\,s_{\tbuy}\le U_t^s.$
\begin{figure}[h]\centering	
	\begin{tikzpicture}[scale=.05]
		\draw[->] (-10,10) -- (20,10);\node at (-5,15){$s_{t-1}$};
		\draw[->] (25,5) -- (25,-15);\node at (25,10) {$t$};\node at (38,-10){$y_{t}$};
		\node[draw,circle,minimum size=4mm] at (25,10) {$t$};\draw[->] (30,10) -- (60,10);
		\draw[->] (25,35) -- (25,15);\node at (38,30){$x_{t}$};
		\node at (45,15){$s_t$};
		\draw[->,dashed] (95,0) -- (120,0);
	\end{tikzpicture}\qquad\qquad\begin{tikzpicture}[scale=.05]
		\draw[->] (-10,10) -- (20,10);\node at (5,15){$s_{t-1}$};
		\draw  (25,10) circle (5cm);\node at (25,10) {$t^-$};\draw[->] (30,10) -- (60,10);
		\draw  (65,10) circle (5cm);\node at (65,10) {$t^+$};\draw[->] (70,10) -- (110,10);
		\draw[->] (25,5) -- (25,-15);\node at (38,-10){$y_{t}$};
		\draw[->] (65,35) -- (65,15);\node at (80,30){$x_{t}$};
		\node at (45,15){$s_{\tsell}$};\node at (90,15){$s_{\tbuy}=s_t $};
	\end{tikzpicture}
	\caption{Splitting the time periods}
\end{figure}

We denote the split periods, ordered chronologically, by
$$\Tspl=\Big(0, 1^{-}, 1^{+}, 2^{-}, 2^{+}, \ldots, \Tsell,\Tbuy\Big)$$
and for $\tau\in\Tspl$ we  define $U_\tau$ to be $\Ust$ when $\tau \in\{\tsell,\tbuy\}$.
For feasible points in $\Pone$, we next identify  periods $\tau\in \Tspl$ when stock levels are at their bounds.

\begin{definition}
	Let $p=(x,y,s)\in\Pone$. We define the sequence of \textbf{stock-extreme periods} of $p$ by
	$$	ET(p)=\big(\tau_0=0,\tau_1,\ldots,\tau_m\big),$$
	where the periods are ordered chronologically, and $\tau\in	ET(p)$ if and only if  $\tau=0$ or
	$	s_{\tau}\in\{0,U_{\tau}\}.	$

	The intervals
	\[	(\tau_{i-1},\tau_i],\qquad i=1,\ldots,m,	\]
	are called the \textbf{stock-extreme intervals} of $p$, and $	(\tau_m,\Tbuy]$	is called the \textbf{last interval} of $p$.
	
	\noindent Finally, we define the set of \textbf{sub-extreme periods} of $p$ to be the following set
\[
ST(p) = \big\{\tsell \in \Tspl\;|\; 0<\yt<\Uyt\} \cup \{\tbuy \in \Tspl\;|\; 0<\xt<\Uxt\big\}
\]
\end{definition}

Note that a split period  $\tau\in\Tspl$ may belong to both $ET(p)$ and $ST(p)$. We are interested in feasible points for which every stock-extreme interval contains at most one sub-extreme period, while the last interval contains none.

\begin{definition}
	A point $p\in\Pone$ is called a \textbf{critical point} if $| ST(p)\cap I|\leq 1$ for any stock-extreme interval $I$ of $p$ and $ST(p)\cap J = \emptyset$ for the last interval $J$ of $p$.
\end{definition}

\noindent We are now ready to characterize the extreme points of $\Pone$.

\begin{theorem}\label{thrm:extsolns}
	A point \(p\) is an extreme point of \(\Pone\) if and only if it is a	critical point. 
\end{theorem}

\begin{proof}
	Let  $p = (x,y,s)$ be a critical point of  $\Pone$. We first show that it is an extreme point. Suppose not, then there exist distinct  points $p^1,p^2\in\Pone$ such that $p=1/2p^1+1/2p^2$.  Note that at any time $t$ such that $\yt\in\{0,\Uyt\}$, we must have $\yt^i=\yt$ for  $i=1,2$. Similarly, for any  $t\in\tset$ such that $\xt\in\{0,\Uxt\}$ we must have $\xt^i = \xt $ for  $i=1,2$.
	In addition if $\stsell = 0~(i.e., s_{t-1}=y_t)$ or $\stbuy = \Ust~(i.e., s_{t}=\Ust),$ we must have $\stsell^i = \stsell$ or $\stbuy^i = \stbuy$, respectively, for  $i=1,2$.	
	Now, since the last interval $J$ of $p$ does not contain any sub-extreme periods, we can conclude that for any period $\tsell$ or $\tbuy$ in the last interval $J$ of $p$, we must have $\yt=\yt^i$ or $\xt=\xt^i$  for $i=1,2$. Also, in any stock-extreme interval $I = (a,b]$ of $p$, there can be at most one period that is sub-extreme. Hence there is at most one period $\tsell$ (or $\tbuy$) where $\yt^i\neq \yt$ (or $\xt^i\neq\xt$). Notice that we must have $\s_a^i = \s_a$ and $\s_b^i = \s_b$ for $i=1,2$ since $a$ and $b$ are stock-extreme periods. Furthermore, $s_b^i = s_a^i - \sum_{t:\tsell \in I} \yt^i + \sum_{t:\tbuy\in I}\xt^i$. Now, since the quantities in this expression agree with $s_a,s_b,\yt$ and $\xt$ in all but one variable, it must be the case that the quantities in this expression agree with $s_a,s_b,\yt$ and $\xt$ in all the variables. Hence we have shown that $p^i = p$ for $i=1,2$. This is a contradiction.
	
	We next argue that if $p=(x,y,s)$ is an extreme point of $\Pone$ then it is a critical point. Suppose $p$ is not a critical point of $\Pone$, then one of the following cases occur:
	
	\textit{Case 1:} Suppose the last interval $J = (a,b]$ of $p$ contains a sub-extreme period $t_m^{k_m}$. Recall that $J$ does not contain any stock-extreme period. Therefore for any period $t^{k}$ in $J$, we must have $0<\s_{t^{k}}<\Ust$. {But then, using a small enough $\epsilon$ we can construct two feasible points by simultaneously increasing (or decreasing) $x_{t_m}$ and $y_{t_m}$ depending on whether $k_m = \textit{buy}$ or $k_m = \textit{sell}$.}  This gives a contradiction.
	
	\textit{Case 2:} Suppose there is a stock-extreme interval $I = (a,b]$ of $p$ containing two sub-extreme periods $t_1^{k_1}$ and $t_2^{k_2}$ with $t_1\leq t_2$. Also assume that $k_1 = \textit{buy} = k_2$. The other three sub-cases are handled similarly. Recall that $I$ does not contain any stock-extreme periods other than time $b$. This means that for any period $t^k$ in the interval $[t_1^{k_1},t_2^{k_2})$, we must have $0<\s_{t^{k}}<\Ust$. {But then, using a small enough $\epsilon$ we can construct two feasible points by simultaneously increasing  $x_{t_1}$ and decreasing $x_{t_2}$, or, vice versa. This gives a contradiction.}
\end{proof}

\subsection{Potential purchase, storage, and sales quantities for extreme points}\label{subsec:potentialstock}

We next describe a set of potential stock levels that can occur in an extreme point of $\Pone$. Additionally, we give an explicit description of potential purchasing and sales quantities in any extreme point of $\Pone$ during time period $t$ when  $s_{t-1}$ and $s_t$  are known. Let
\begin{align}\label{def:St}
	S_t =&   \Big\{\big\{K-\sum_{i=1}^t v_i \Uyi + \sum_{i=1}^t u_i\Uxi\::\:K\in \{0,\snought,\Us_1,\ldots,\Ust\}, v,u\in\{0,1\}^t\big\}\\ 
	&\quad\bigcup 
	\big\{K' + \sum_{i=t+1}^T v'_i\Uyi - \sum_{i=t+1}^Tu'_i\Uxi\::\: K'\in \{0,\Us_{t+1},\ldots,\Us_T\},~ v',u'\in\{0,1\}^{T-t}\big\}\Big\}\bigcap \Big[0,\Ust\Big]\nonumber
\end{align}
\begin{theorem}\label{thrm:stock}
If a point \(p=(x,y,s)\) is an extreme point of \(\Pone\) then $s_t\in S_t$ for all $t\in\tset$.
\end{theorem}

\begin{proof}
	Since every extreme point of the polytope $\Pone$ is also a critical point of $\Pone$, we know that the stock level at the end of any time period $t$ is obtained in one of two ways. Either we start from the last stock-extreme time $t'\leq t$ and only purchase and sell at full or zero capacity till time period $t$. In this case $s_t = K-\sum_{i=1}^t v_i \Uyt + \sum_{i=1}^t u_i\Uxt$ where $K\in \{0,\snought,\Us_1,\ldots,\Ust\}$ and $u_i,v_i \in \{0,1\}$. Otherwise, there exists a sub-extreme time $t_1$ such that $t'< t_1\leq t$. But then, starting at time $t_1$, we can only purchase and sell at full or zero capacity until the next extreme time $t_2\geq t$. In this case $s_t = K' + \sum_{i=t+1}^T v'_i\Uyt - \sum_{i=t+1}^Tu'_i\Uxt$ where $K'\in \{0,\Ust,\Us_{t+1},\ldots,\Us_T\}$ and $u'_i,v'_i \in \{0,1\}$.
\end{proof}

\begin{remark}\label{remark1}The description of $S_t$ above can be tightened as it does not take into account the feasibility of the sequence of purchase and sale quantities. This, however, does not lead to an improvement in the size complexity of the set $S_t$. Let 
	\begin{equation}\label{def:potstockbound}
		\potstockbound = \max_{t\in\tset}\lnorm S_t\rnorm
	\end{equation}
	The size of $S$ is bounded by $O(T4^T)$ in the worst case. In addition when the bounds are integral, $S\leq \max_{t\in\tset}\{\Ust\}$. Note that rational data can be made integral by scaling.\end{remark}

We next characterize the quantities $\xt$ and $\yt$ in extreme points of $\Pone$ when moving from a stock level $\s_{t-1}$ to a stock level $\st$. 
Let $\st-\st{} _{-1} = \delta$, then $\xt$ and $\yt$ must satisfy
$$\Uxt\ge\xt\ge0,\qquad\Uyt\ge\yt\ge0,\qquad\st{}_{-1}\ge\yt,\qquad\xt-\yt=\delta.$$
Observe that any particular values of $\yt$ and $\xt$ creates a \textit{sell-buy cycle} of size $\min\{\yt,\xt\}$. The minimum size of this sell-buy cycle is equal to zero. In this case,
\begin{align*}
	\yt = (\s_{t-1}-\st)^+,\;\;\;\;\xt = (\st-\s_{t-1})^+\tag{a}\label{a}
\end{align*}
And the maximum size occurs either when \(\xt = \Uxt\) or when
\(\yt = \min\{\s_{t-1},\Uyt\}\). In this case,
\begin{align*}
	\yt = (\s_{t-1}-\st)^+ + \Delta_t(\s_{t-1},\st),
	\qquad
	\xt = (\st-\s_{t-1})^+ + \Delta_t(\s_{t-1},\st),
	\tag{b}\label{b}
\end{align*}
where \(\Delta_t(\s_{t-1},\st)\) denotes the maximum size of a sell-buy cycle
when the stock level moves from \(\s_{t-1}\) to \(\st\) during period \(t\). In
particular,
\begin{align*}
\Delta_t(\s_{t-1},\st) =
	\begin{cases}
		\min\{\,\s_{t-1},\,\Uyt,\,\Uxt-(\st-\s_{t-1})\,\}, & \text{if } \st\ge \s_{t-1},\\[2mm]
		\min\{\,\st\,,\Uxt\,,\Uyt-(\s_{t-1}-\st)\,\}, & \text{if } \st\le \s_{t-1}.
	\end{cases}
\end{align*}

At any extreme point of $\Pone$, the size of the sell-buy cycle created by the values $\yt$ and $\xt$ must be either zero or maximum. If not, we can add and subtract a small quantity $\epsilon$ from both $\yt$ and $\xt$ and still remain feasible. Thus for any extreme point, the purchase and sale quantities are given by either  \eqref{a} or \eqref{b}.

\section{Fixed Costs}\label{sec:fixedcosts}

We next turn our attention to Problem (\ref{prob:2}) and the  sets $\Qtwo$ and $\Ptwo$. We first characterize the extreme points of $\Ptwo$. We then give an algorithm to solve \eqref{prob:2} and an extended LP formulation for $\Ptwo$. 
We start with the following definition:

\begin{definition}\label{def:extension}
	For any $p \in \Pone$, let the set of \textit{extensions} of $p$ to be 
	$$Q(p) = \{(\x,\y,\s,\w,\z)\in \Qtwo\::\: \big(\x,\y,\s)=p\big\}.$$
	Let $Q^*(p)$ be the unique point in $Q(p)$ that satisfies $\wt = 1$ iff $\xt>0$ and $\zt=1$ iff $\yt>0$ for all $t\in\tset.$ 
	Note that under the assumption $\htt,\gt\le0$, the point $Q^*(p)$ has maximum objective value among all points in $Q(p)$.
	
\end{definition}

The next lemma relates the extreme points of the polytope $\Ptwo$ to the extreme points of the polytope $\Pone$. Then, using the characterization of extreme points of $\Pone$ given in Theorem \ref{thrm:extsolns}, we will obtain a characterization of extreme points of the polytope $\Ptwo$.

\begin{lemma}\label{lem:fixedcosts}
	Let $(\Bar{x},\Bar{y},\Bar{s},\Bar{w},\Bar{z})$ be an extreme point of $\Ptwo$, then $(\Bar{x},\Bar{y},\Bar{s})$ is an extreme point of $\Pone$. Conversely, if $p=(\Bar{x},\Bar{y},\Bar{s})$ is an extreme point of $\Pone$, then every point in $Q(p)$ is an extreme point of $\Ptwo$.
\end{lemma}

\begin{proof}
	
Let \((\Bar{x},\Bar{y},\Bar{s},\Bar{w},\Bar{z})\) be an extreme point of \(\Ptwo\). Since \(\Bar{w},\Bar{z}\in\{0,1\}^T\), the constraints \(0\le x_t\le \Uxt \wt\) and \(0\le y_t\le \Uyt \zt\) imply that \(\Bar{x}_t=0\) whenever \(\Bar{w}_t=0\) and \(\Bar{y}_t=0\) whenever \(\Bar{z}_t=0\). Thus \((\Bar{x},\Bar{y},\Bar{s})\) lies in the face
\[F=\{(x,y,s)\in\Pone : x_t=0 \ \forall\, t \text{ with } \Bar{w}_t=0,\; y_t=0 \ \forall\, t \text{ with } \Bar{z}_t=0\}.\]
Moreover, every point in \(F\) can be extended with the same \((\Bar{w},\Bar{z})\) to a feasible point of \(\Ptwo\). Therefore \((\Bar{x},\Bar{y},\Bar{s})\) must be extreme in \(F\), and hence extreme in \(\Pone\).

For the reverse direction, let \(p=(\Bar{x},\Bar{y},\Bar{s})\) be an extreme point of \(\Pone\), and let \(q\in Q(p)\). 
Suppose that $	q=\tfrac12 q^1+\tfrac12 q^2$ for some \(q^1,q^2\in\Ptwo\). 
Projecting onto the \((x,y,s)\)-coordinates gives $	p=\tfrac12 p^1+\tfrac12 p^2,$ where \(p^i=\proj_{x,y,s}(q^i)\in\Pone\) for \(i=1,2\). Since \(p\) is extreme, we must have \(p^1=p^2=p\). Therefore, \(q^1\) and \(q^2\) agree with \(q\) in the \((x,y,s)\)-coordinates. Because \(q\in Q(p)\), its \((w,z)\)-coordinates are binary, and since each coordinate of \(q\) is the average of the corresponding coordinates of \(q^1\) and \(q^2\), which lie in \([0,1]\), those coordinates must also agree. Hence \(q^1=q^2=q\), and so \(q\) is an extreme point of \(\Ptwo\).
\end{proof}

\begin{corollary}\label{cor:extP2}
	The set of extreme points of $\Ptwo$ is equal to  $\bigcup_{p\in \mathcal{C}(\Pone)} Q(p)$ where $\mathcal{C}(\Pone)$ are the critical points of the polytope $\Pone$. Consequently, $\proj_{x,y,s}(\Ptwo) = \Pone$.
\end{corollary}

\subsection{Network Flow Formulation for Problem (\ref{prob:2})}\label{subsec:networkP2}

In this subsection, we will construct a  directed acyclic network $G_1 = (V,A)$ with parallel edges and reformulate Problem (\ref{prob:2}) as a longest path (or, minimum cost flow) problem in this network. Using the set of potentially optimal stock levels identified in Corollary \ref{cor:extP2} and equation \eqref{def:St}, the network will allow us to move between these stock levels according to equations (a) and (b). The node set $V := \bigcup_{0}^T V_t$ where $V_t=S_t$ and $S_0 = \{\snought\}$. The edge set $A = \bigcup_{1}^T A_t$ where $A_t$ contains two parallel edges directed from $s\in V_{t-1}$ to $s'\in V_t$ if $s'\in [s-\Uyt,s+\Uxt]$. 
These two edges correspond to the two possible sell-buy cycles described by equations (a) and (b). Note that $G_1$ has the following properties:
\begin{enumerate}
	\item Any path from $V_0$ to $V_T$ corresponds to a feasible point $(\x,\y,\s) \in \Pone$.
	\item Every critical point $(\x,\y,\s)$ of $\Pone$ corresponds to a path from $V_0$ to $V_T$.
\end{enumerate}

For any edge $a\in A_t$, the values of $\xt$ and $\yt$ are fixed and so we can set $\wt = 1$ iff $\xt>0$ and $\zt=1$ iff $\yt>0$ and set the weight of edge $a$ to be the corresponding pay-off at time $t$, i.e. 
$$\pt\yt+\ct\xt+\rt\st+\htt\zt+\gt\wt.$$ 
Then, any path corresponding to a feasible point $p\in \Pone$ will have total weight $Obj(Q^*(p))$ where $Obj$ is the objective function of problem (\ref{prob:2}). Thus, by Corollary \ref{cor:extP2} and Property 1 and 2 above, a longest path from $V_0$ to a node in $V_T$ in $G_1$ corresponds to an optimal solution of Problem \eqref{prob:2}.

\medskip\noindent\textbf{Algorithm 1:}\label{subsec:alg1}
We can therefore solve problem \eqref{prob:2} by finding the longest path from $V_0$ to $V_T$ in $G_1$. Note that while constructing the network, we can use Remark \ref{remark1} and construct the sets $V_t$ accordingly. Then $|V|\le TS$  and $|A|\le 2TS^2$ and hence  \textit{Algorithm 1} runs in time $O(TS^2)$ which is pseudo-polynomial if the bounds are rational.

\subsection{Extended Linear Formulation for $\Ptwo$}\label{subsec:extQ2}

We next provide an extended LP formulation for the polytope $\Ptwo$ by relating the variables $\xt,\yt,\st,\wt,\zt$ to the network flow variables of network $G_1$. Let $\alpha_a$ be the flow variable corresponding to edge $a\in A$. For any edge $a\in A$, define $q^x(a)$ to be the value of $\xt$ corresponding to edge $a$ and similarly define $q^y(a)$ to be the value of $\yt$. Then the network flow constraints are,
\begin{align*}
	\sum_{a\in \delta^{in}(v)} \alpha_a &= \sum_{a\in \delta^{out}(v)} \alpha_a &&\forall v\in \bigcup_{t=1}^{T-1} V_t&&\tag{i}\\
	\sum_{a\in\delta^{out}(\snought)}\alpha_a &= 1\tag{ii}\\
	\alpha_a &\geq 0 && \forall a\in A\tag{iii}
\end{align*}
Additionally, we can relate the variables $\xt,\yt,\st,\wt,\zt$ to the network flow variables $\alpha_a$ as follows,
\begin{align*}
	x_t &= \sum_{a\in A_t} q^x(a) \alpha_a\qquad&&y_t = \sum_{a\in A_t} q^y(a) \alpha_a &&\forall\;t\in\tset&&\tag{iv}\\
	s_t &= s_{t-1} -\yt +\xt &&&&\forall\;t\in\tset\tag{v}\\[.1cm]
	\wt &\geq \sum_{a\in A_t : q^x(a)>0}\alpha_a\qquad	&&\zt \geq \sum_{a\in A_t : q^y(a)>0} \alpha_a\ &&\forall\;t\in\tset\tag{vi}\\
	\wt&\leq 1\qquad && \zt\leq 1&&\forall\;t\in\tset\tag{vii}
\end{align*}

Let $\Ptwo'$ be the polytope defined by the constraints (i)-(vii). We will show that the polytope $\Ptwo'$ is an extended linear formulation for the polytope $\Ptwo$ using the following intermediate lemma.

\begin{lemma}\label{lem:integralityP2'}
	Let $q'=(\Bar{\alpha},\Bar{x},\Bar{y},\Bar{w},\Bar{z},\Bar{s})$ be an extreme point of $\Ptwo'$. Then, $\Bar{\alpha},\Bar{w}$ and $\Bar{z}$ are integral.
\end{lemma}

\begin{proof}
	Suppose for the sake of contradiction that $(\Bar{\alpha})$ is not integral. Observe that $(\Bar{\alpha})$ are feasible to the unit flow LP described by equations (i)-(iii). We know that the unit flow LP is integral and thus, there exist feasible points $\alpha^1$ and $\alpha^2$ distinct from $(\Bar{\alpha})$ such that $(\Bar{\alpha}) = 1/2(\alpha^1+\alpha^2)$. We now extend each vector $\alpha^i$ to feasible points in $\Ptwo'$. The $x^i,y^i$ and $s^i$ variables are forced by equations (iv)-(v) and we choose $w^i$ and $z^i$ so that inequalities (vi) are tight equalities. Thus, we obtain points $p^i=(\alpha^i,x^i,y^i,w^i,z^i,s^i)$ that are feasible points in $\Ptwo'$. Let $w^*$ and $z^*$ be the vectors obtained using $(\Bar{\alpha})$ when we set inequalities (vi) to tight equalities. Then, we observe that the point $p=(\Bar{\alpha},\Bar{x},\Bar{y},w^*,z^*,\Bar{s})$ lies in the convex hull of the points $p^1$ and $p^2$. Furthermore, $1\geq \Bar{w}\geq w^*$ and $1\geq\Bar{z}\geq z^*$ and we can replace $w^*_t$ or $\zt^*$ by $1$ in the point $p$ to still remain feasible in $\Ptwo'$. But then, we can express the vector $q$ as a convex combination of points distinct from $q'$ that are feasible in $\Ptwo'$. This is a contradiction since the point $q'$ was assumed to be an extreme point.
	
	Thus $\Bar{\alpha}$ is proven to be integral. At any extreme point of $\Ptwo'$, one of the inequalities containing $\wt$ must be tight. $\wt$ occurs in exactly two inequalities described by (vi) and (vii). Either of these being tight would imply that $\wt$ is an integer. A similar argument holds for $\zt$ and the lemma is proved.
\end{proof}

\begin{theorem}
	The polytope $\Ptwo'$ is an extended linear formulation for the polytope $\Ptwo$.
\end{theorem}

\begin{proof}
	We will show that every extreme point of $\Ptwo'$ is an extension of a feasible point of $\Ptwo$ and also every extreme point of $\Ptwo$ can be extended to a feasible point in $\Ptwo'$.
	
	Let $q'=(\Bar{\alpha},\Bar{x},\Bar{y},\Bar{w},\Bar{z},\Bar{s})$ be an extreme point of $\Ptwo'$. Then the lemma above proves that $(\Bar{\alpha},\Bar{w},\Bar{z})$ is integral. Since $\Bar{\alpha}$ is integral, it is the incidence vector of a path in the network  and hence  $(\Bar{x},\Bar{y},\Bar{s})$ is feasible for $\Pone$.  Furthermore, inequalities (vi) ensure that $\Bar{z}_t=1$ if $\Bar{y}_t>0$ and $\Bar{w}_t=1$ if $\Bar{x}_t>0$. Hence $(\Bar{x},\Bar{y},\Bar{w},\Bar{z},\Bar{s})$ is indeed a feasible point in $\Ptwo$.
	
	Now, let $q = (\Bar{x},\Bar{y},\Bar{w},\Bar{z},\Bar{s})$ be an extreme point of $\Ptwo$. Then, we know from Corollary \ref{cor:extP2} that $p = (\Bar{x},\Bar{y},\Bar{s})$ is an extreme point of $\Pone$. But then, from the construction of the network $G_1$, we know that the point $p$ corresponds to a path of the network and we can set $(\Bar{\alpha})$ to be the identity vector of the path corresponding to the point $p$. Doing so will ensure that the point $q'=(\Bar{\alpha},\Bar{x},\Bar{y},\Bar{w},\Bar{z},\Bar{s})$ satisfies equations (i)-(vii) and so the point $q$ can be extended to a feasible point in $\Ptwo'$. This completes the proof of the theorem.
\end{proof}

The extended formulation $\Ptwo'$ has $O(TS^2)$ variables and $O(TS^2)$ constraints where $S$ is defined in equation \eqref{def:potstockbound}.

\section{Complementarity Constraints}\label{sec:comp}

In this section, we turn our attention to Problem (\ref{prob:3}) and the corresponding sets $\Qthree$ and $\Pthree = \conv(\Qthree)$. Recall the definition of $Q(p)$ provided in Definition \ref{def:extension} and let $\mathcal{C}'(\Pone)$ be the set of critical points of $\Pone$ such that $\xt\yt = 0$ for all $t\in\tset$. Then,

\begin{lemma}\label{lem:comp}
	Let $(\Bar{x},\Bar{y},\Bar{s},\Bar{w},\Bar{z})$ be an extreme point of $\Pthree$, then $(\Bar{x},\Bar{y},\Bar{s}) \in \mathcal{C}'(\Pone) $. Conversely, if $p=(\Bar{x},\Bar{y},\Bar{s}) \in \mathcal{C}'(\Pone)$, then every point in $Q(p)$ is an extreme point of $\Pthree$.
\end{lemma}

\begin{proof}
	Let $q = (\Bar{x},\Bar{y},\Bar{s},\Bar{w},\Bar{z})$ be an extreme point of $\Pthree$. Clearly  point $p=(\Bar{x},\Bar{y},\Bar{s})$ is feasible for $\Pone$. If $p$ is not an extreme point of $\Pone$, then there exist two distinct points $p^1,p^2\in\Pone$ such that $p=\tfrac12(p^1+p^2)$. 
	Since $q\in\Pthree$, we have $\Bar{x}_t\Bar{y}_t=0$ for all $t\in\tset$. Because all variables are nonnegative,  whenever $\Bar{x}_t=0$ we must have $x_t^1=x_t^2=0$, and whenever $\Bar{y}_t=0$ we must have $y_t^1=y_t^2=0$. Hence the complementarity constraints $x_t y_t=0$ remain satisfied by both $p^1$ and $p^2$. Moreover, if $\Bar{w}_t=0$ then $\Bar{x}_t=0$, and if $\Bar{z}_t=0$ then $\Bar{y}_t=0$, so the linking constraints with $\Bar{w}$ and $\Bar{z}$ are also preserved. Therefore the points $q^i=(p^i,\Bar{w},\Bar{z})$ are feasible in $\Pthree$ for $i=1,2$, and $q=\tfrac12(q^1+q^2)$, contradicting the assumption that $q$ is an extreme point of $\Pthree$.
	
	For the reverse direction, let $p = (\Bar{x},\Bar{y},\Bar{s})\in \mathcal{C}'(\Pone)$ and let 
	$q = (\Bar{x},\Bar{y},\Bar{s}, \Bar{w},\Bar{z}) \in Q(p)$. Note that $\Bar{w}$ and $\Bar{z}$ are binary.
	If $q=1/2(q^1+q^2)$ where $q^1,q^2 \in \Pthree$, then $\proj_{w,z}(q) = \proj_{w,z}(q^1) = \proj_{w,z}(q^2)$. Setting $p^i = \proj_{x,y,s}(q^i)$, we  obtain that $p = 1/2(p^1+p^2)$, a contradiction.\qed
\end{proof}

\begin{corollary}\label{cor:extP3}
	The set of extreme points of $\Pthree$ is equal to the set $\bigcup_{p\in \mathcal{C}'(\Pone)} Q(p)$.
\end{corollary}

Using the above corollary, we obtain a network formulation for Problem \eqref{prob:3}  by making a simple modification to the network $G_1$. We know that every critical point of $\Pone$ corresponds to a path in  $G_1$.
Moreover, equations (a) and (b) describe the two possible sell-buy cycles that can occur on each edge of $G_1$. 
Under the complementarity constraints, only the edges corresponding to equation~(a) can appear in the path corresponding to a point $p\in\mathcal{C}'(\Pone)$.
Conversely, every such path corresponds to a feasible point in $\Qthree$. Let $G_2$ be the network obtained from $G_1$ by deleting all edges corresponding to equation (b). Then, the network $G_2$ has the following properties:
  
\begin{enumerate}
	\item Any path from $V_0$ to $V_T$ in $G_2$ corresponds to a feasible point $(x,y,s) \in \Pone$ such that $x_ty_t = 0$ for all $t\in\tset$.
	\item Every critical point $(x,y,s)$ of $\Pone$ such that $x_ty_t = 0$ for all $t\in\tset$ corresponds to a path from $V_0$ to $V_T$ in $G_2$.
\end{enumerate}

\noindent This leads to the following algorithm and extended formulation. 

\medskip\noindent\textbf{Algorithm 2:}\label{subsec:alg2} Problem \eqref{prob:3} can be solved by finding a longest path from $V_0$ to $V_T$ in $G_2$. Similar to algorithm 1, the running time is $O(TS^2)$ which is pseudo-polynomial if the bounds are rational. Recall $S$ is defined in equation \eqref{def:potstockbound} and note that the size of $S$ is bounded by $O(T3^T)$ in the worst case for Problem \eqref{prob:3}.

\medskip\noindent\textbf{Extended  formulation for $\Pthree$:} Network $G_2$ leads to a linear formulation $\Pthree'$ which is an extended formulation of $\Pthree$. In fact, as $G_2$ is obtained from $G_1$ by deleting all edges corresponding to equations (b), $\Pthree'$ is obtained from $\Ptwo'$ by setting $\alpha_a = 0$ for all edges $a\in A$ that correspond to equation (b). This extended formulation has $O(TS^2)$ variables and $O(TS^2)$ constraints.

\section{Polynomially Solvable Special Cases}\label{sec:special}

In this section, we  consider some special cases of the Problems discussed in this paper. We will assume a structure on the time-varying bounds $\Uxt$, $\Uyt$ and $\Ust$ and analyze the running times of Algorithm 1 and Algorithm 2. We will also analyze the size of the extended linear formulations for polytopes $\Ptwo$ and $\Pthree$. Recall the definition of the set $S_t$ provided in equation \eqref{def:St} and the definition of the quantity $S$ provided in equation \eqref{def:potstockbound}

We first consider the special case where $\Ust = \Uxt = \Uyt = U$, and observe that the set $S_t$ is always a subset of the set $\{0,\snought,U\}$, and therefore, $S=3$. Consequently, Algorithm 1 and Algorithms 2 both have a running time of $O(T)$ to solve problems (\ref{prob:2}) and  (\ref{prob:3}). In addition, the extended linear formulations for polytopes $\Ptwo$ and $\Pthree$ are of size $O(T)$. This case was recently studied by Wolsey and Yaman~\cite{wolsey} where they provide algorithms with running time $O(T)$ to solve  \eqref{prob:2} and  \eqref{prob:3}. They also provide extended linear formulations for polytopes $\Ptwo$ and $\Pthree$ of size $O(T)$. The ideas for our algorithms and extended formulations are borrowed from their work but made to work in a more general framework. 




\subsection{Time-independent bounds}

We next consider the case where the bounds on purchase, storage and sales are not time-varying. Thus assume that $\Ust = \Us, \Uxt = \Ux$ and $\Uyt = \Uy$ for all times $t\in\tset$. 
Note that in this case, we can also assume that $ \Us\ge\max\{\Uy,\Ux\}$ as sales occur before any purchases are made.
Prior to our work, there were no polynomial time algorithms known for problems \eqref{prob:2} and \eqref{prob:3} in this setting. We bound the size of $S$ in this setting using the following lemma.

\begin{lemma}
	If $\Ust = \Us, \Uxt = \Ux$ and $\Uyt = \Uy$, then $S\leq 4T(\gamma+1)$ where  $\gamma = \min\big\{T,\frac{\Us}{\max\{\Uy,\Ux\}}\big\}$.
\end{lemma}

\begin{proof}
	Using definition \eqref{def:St}, we see that the set $S_t$ can be described by
	\begin{align*}
		S_t \subseteq& \Bigg\{\Big\{K-k\Uy + k'\Ux\::\:K\in \{0,\snought,\Us\},~ k,k'\in\{0,1,\ldots,t\}\Big\}\\ 
		&\quad\bigcup 
		\Big\{K' + k\Uy - k'\Ux\::\:K'\in \{0,\Us\},~ k,k'\in\{0,1,\ldots,T-t\}\bigg\}\Bigg\}\bigcap \bigg[0,\Us\bigg]
	\end{align*}
	Now assume $\Ux\geq \Uy$. The other case is handled similarly. For each value of $K$ or $K'$, we will try to bound the number of permissible pairs $(k,k')$. When $K=0$, we must have $0\leq -k\Uy+k'\Ux\leq \Us$. Thus, $k\Uy/\Ux\leq k'\leq (k\Uy+\Us)/(\Ux)$. But then, for any fixed $k$, there are at most $\min\{t,{\Us}/{\Ux}+1\}$ permissible values for $k'$. Also, there are at most $t$ permissible values for $k$. This implies that there are at most $t(\gamma+1)$ permissible values for the pair $(k,k')$ when $K=0$. When $K = \Us$, we must have $0\leq k\Uy-k'\Ux\leq \Us$. Thus, ${(k\Uy-\Us)}/{\Ux}\leq k'\leq {k\Uy}/{\Ux}$. Again, for any fixed $k$, there are at most $\min\{t,{\Us}/{\Ux}+1\}$ permissible values for $k'$. This implies that there are at most $t(\gamma+1)$ permissible values for the pair $(k,k')$ when $K=\Us$. Next, we can take the union of the above two sets of permissible pairs $(k,k')$ as a superset of the permissible pairs $(k,k')$ when $K=\snought$. Finally, the analysis for $K' = 0$ or $K'=\Us$ is exactly the same, replacing $t$ by $T-t$. Hence $\lnorm S_t\rnorm\leq 4t(\gamma+1)+2(T-t)(\gamma+1)\leq 4T(\gamma+1)$.
\end{proof}

\begin{corollary}
	An immediate corollary of the above lemma is that Algorithm 1 and Algorithm 2 both have running times $O(T^3\gamma^2)$ in the setting where $\Ust=\Us,\Uxt=\Ux$ and $\Uyt=\Uy$ for all $t\in\tset$. Furthermore, the extended linear formulations for polytopes $\Ptwo$ and $\Pthree$ are of size $O(T^3\gamma^2)$.
\end{corollary}

\subsection{Fixed-dimensional lattice structure  on time-varying bounds}

We next  consider the case where all time-varying bounds and the initial stock lie on the lattice generated by some fixed positive constants $d_1,\ldots,d_k$. Thus,
\[\Ust,\Uxt,\Uyt,\snought \in \Big\{\sum_{i=1}^{k}\alpha_i d_i\::\: \alpha_i\in \mathbb{Z}_+\Big\}.\]

For each $i=1,\ldots,k$, let $r_i$ denote an upper bound on the maximum value that $\alpha_i$ can take in the representation above.
For example, we can take $r_i=\lfloor \bar U/d_i\rfloor$ where 
$$\bar U=\max\big\{s_0,\,\max_{t\in\tset}U_t^s\,,\max_{t\in\tset}U_t^x,\,\max_{t\in\tset}U_t^y\big\}.$$
Using definition \eqref{def:St}, the set $S_t$ satisfies
\begin{align*}
	S_t \subseteq\Big\{\sum_{i=1}^k\beta_i d_i\::\: \beta_i\in[-Tr_i,(T+1)r_i]\cap\mathbb{Z}\Big\}.
\end{align*}
Hence,
	$$	\lnorm S_t\rnorm \le \prod_{i=1}^k \big((2T+1)r_i+1\big),	\quad
	\text{and therefore	}\quad
	S \le \prod_{i=1}^k \big((2T+1)r_i+1\big),	$$
where $S=\max_{t\in\tset} |S_t|$, as defined earlier.

In particular, if the lattice dimension $k$ is fixed, then $S=O(T^k)$ and consequently Algorithms~1 and~2 run in polynomial	time.  
Moreover, the extended formulations for $\Ptwo$ and $\Pthree$ have  polynomial size.

As an application of this setting, consider a scenario in which certain sections of the warehouse are subject to scheduled maintenance. 
For example suppose that at each time period $t$, the warehouse capacity takes one of three values \textit{high, medium} or \textit{low} represented by $d_1,d_2,d_3$ depending on which sections of the warehouse are under maintenance. If the bounds on purchase and sales also take on values in $\{d_1,d_2,d_3\}$  at any time $t$, then $k=3$, and our algorithms have running time $O(T^7)$. If the bounds on purchase and sales are not time-varying but could take some other values, then $k=5$ and the above analysis shows that our algorithms have running time $O(T^{11})$.
More generally, if the total number of distinct values taken by the storage, purchase, and sales bounds is bounded by a constant, then our algorithms run in polynomial time.

\section{Extensions}\label{sec:further}

In this section, we discuss several extensions and variants of the warehouse problem that can be handled by our methods. Each extension introduces optional modifications to Problems~\eqref{prob:1}, \eqref{prob:2}, and \eqref{prob:3}. These extensions do not change the underlying polyhedral framework developed in Sections~\ref{sec:extreme}--\ref{sec:comp}; they only require appropriate modifications to the characterization of extreme points, the network representation, or the edge weights. Moreover, these extensions can be combined while preserving our overall solution approach.
\subsection{Fixed costs for storage}

We next consider introducing per-unit and fixed costs for storage. This can be modeled by introducing binary variables $v_t$ such that $v_t = 1$ if $s_t>0$. This leads to the following formulation:

\begin{align*}\label{prob:4}
	\max & \;\;\;\;\sum_{t=1}^\tbound ~(\pt\yt + \ct\xt  +\rt\st +\htt\zt + \gt\wt + f_t v_t) \tag{\bf{IP3}}\\
	\text{s.t.}&\;\;\;\; \st = \s_{t-1} - \yt + \xt & \forall\; t\in\tset\\
	&\;\;\;\;\yt \leq \s_{t-1} &\forall\; t\in\tset \\
	&\;\;\;\; 0\leq \st\leq \Ust v_t,\;\;\;\;0\leq \yt\leq \Uyt\zt,\;\;\;\;0\leq \xt\leq \Uxt\wt& \forall\; t\in\tset\\
	&\;\;\;\; \wt,\zt, v_t\in\{0,1\} &\forall\; t\in\tset
\end{align*} 
Note that in this model we can assume  $f_t<0$ for all $t\in\tset$ without loss of generality.
Let the feasible region of this integer program be called $Q_3$ and its convex hull $P_3 = \conv(Q_3)$. In this setting, for a feasible point $p=(\bar x,\bar y,\bar s) \in \Pone$, we can modify the definition of the set $Q(p)$ given in Definition  $\ref{def:extension}$ to the following 
\[Q'(p) = \big\{(\bar x,\bar y,\bar s,w,z,v) \in Q_3\big\}\]
With this change, we can modify Lemma \ref{lem:fixedcosts} as follows:

\begin{lemma}
	Let $(\Bar{x},\Bar{y},\Bar{s},\Bar{w},\Bar{z},\Bar{v})$ be an extreme point of $P_3$, then $(\Bar{x},\Bar{y},\Bar{s})$ is an extreme point of $\Pone$. Conversely, if $p=(\Bar{x},\Bar{y},\Bar{s})$ is an extreme point of $\Pone$, then every point in $Q'(p)$ is an extreme point of $P_3$.
\end{lemma}

The proof is identical to that of Lemma \ref{lem:fixedcosts} after including the additional binary variables $v$.
We also immediately obtain the following corollary:

\begin{corollary}
	The set of extreme points of $P_3$ is equal to $\cup_{p\in \mathcal{C}(\Pone)} Q'(p)$ where $\mathcal{C}(\Pone)$ denotes the set of critical points of  $\Pone$. Consequently $\proj_{x,y,s}(P_3) = \Pone$.
\end{corollary}

We know that critical points of $\Pone$ correspond to paths in the network $G_1$. Hence, if we change the edge weights in the networks $G_1$ and $G_2$ to reflect the per-unit and fixed costs for storage, then a longest path in the network will still correspond to an optimal solution of the corresponding warehouse problem. For any edge, the value of $s_t$ is fixed and therefore we increase the  weight of that edge by $\ft$ if $s_t>0$. 

Note that one can also include the complementarity constraints $x_ty_t = 0$ for all $t\in \tset$ as the arguments in Section \ref{sec:comp} can be extended. Thus, we still have algorithms to solve the basic problems even when per-unit and fixed costs for storage are introduced. The running times of these algorithms are exactly the same as those of Algorithm~1, both in the general case and in the special cases discussed in Section~\ref{sec:special}.

\subsection{Lower bounds on sales, storage, purchases}

Now consider the case when there are also time-varying lower bounds on sales, storage and purchase quantities given by $L^y_t, L^s_t, L^x_t$. 
This leads to the following formulation:

\begin{align*}\label{prob:5}
	\max & \;\;\;\;\sum_{t=1}^\tbound ~(\pt\yt + \ct\xt  +\rt\st +\htt\zt + \gt\wt  + f_t v_t )\tag{\bf{IP4}}\\
	\text{s.t.}&\;\;\;\; \st = \s_{t-1} - \yt + \xt & \forall\; t\in\tset\\
	&\;\;\;\;\yt \leq \s_{t-1} &\forall\; t\in\tset \\
	&\;\;\;\; L^s_t v_t\leq \st\leq \Ust v_t,\;\;\;\;L^y_t z_t\leq \yt\leq \Uyt\zt,\;\;\;\; L^x_t w_t\leq \xt\leq \Uxt\wt& \forall\; t\in\tset\\
	&\;\;\;\; \wt,\zt, v_t\in\{0,1\} &\forall\; t\in\tset
\end{align*}

Let the feasible region of the above integer program be $Q_4$ and its convex hull be $P_4 = \conv(Q_4)$. Consider the basic version of this problem without fixed costs i.e. Problem \eqref{prob:1} with lower bounds added on. Let the feasible region of the polytope here be called $\Pone'$. In this case, we will need to modify the characterization of critical points of $\Pone'$ as follows: For any point $p$ in $\Pone'$, define its \textit{stock-extreme times} to be split time periods at which $s_{t^k} \in \{L^s_t, U^s_t\}$. Define its \textit{sub-extreme times} to be times $t^{\text{sell}}$ at which $L^y_t < \yt < U^y_t$ or times $t^{\text{buy}}$ at which $L^x_t < \xt <U^x_t$. With these changes, we still define critical points of the polytope $\Pone'$ to be points that have  at most one sub-extreme time in any stock-extreme interval and no sub-extreme times in its last interval. Then, the proof of Theorem \ref{thrm:extsolns} carries over verbatim after replacing the lower bound of 0 with the corresponding lower bounds throughout. Hence, critical points of $\Pone'$ are exactly the extreme points of $\Pone'$. Consequently, the stock level at the end of  time period $t$ in an extreme point of $\Pone'$ will be described by either starting at a stock-extreme time $t'\leq t$ and purchasing and selling at upper or lower bounds till time $t$ or by starting at time $t$ and purchasing and selling at their upper or lower bounds until a stock-extreme time $t' \geq t$. Hence we get a new characterization of the set of potential stock levels at the end of time $t$ in any extreme point of $\Pone'$.

\begin{align*}
	S'_t =&   \Big\{\big\{K-\sum_{i=1}^t (v^1_i \Uy_i + v^2_i L^y_i) + \sum_{i=1}^t (u^1_i\Ux_i + u^2_i L^x_i)\::\:K\in \{L^s_1,\ldots,L^s_t,\snought,\Us_1,\ldots,\Ust\},\\ 
	& \quad\quad v^j,u^j\in\{0,1\}^t,\quad v^1_i+v^2_i = u^1_i+u^2_i = 1\big\}\bigcup\\ 
	& \big\{K' + \sum_{i=t+1}^T (v^{1'}_i\Uy_i + v^{2'}_i L^y_i) - \sum_{i=t+1}^T(u^{1'}_i\Ux_i + u^{2'}_i L^x_i)\::\: K'\in \{L^s_{t+1},\ldots,L^s_T,\Us_{t+1},\ldots,\Us_T\},\\ 
	& \quad\quad v^{j'},u^{j'}\in\{0,1\}^{T-t},\quad v^{1'}_i+v^{2'}_i = u^{1'}_i+u^{2'}_i = 1\big\}\Big\}\bigcap \Big[L^s_t,\Ust\Big]\nonumber
\end{align*}
Once the values of $s_{t-1}$ and $s_t$ are fixed, the quantities $x_t$ and $y_t$ at an extreme point of $\Pone'$ must correspond to either the minimum or the maximum feasible sell-buy cycle. Consequently, there are only two possible choices for $(x_t,y_t)$.
Furthermore, any extreme point of $P_4$ can be shown to be an extension of an extreme point of $\Pone'$ analogous to Lemma \ref{lem:fixedcosts}. These results lead directly to a network flow formulation for Problem~\eqref{prob:5}, yielding the same algorithmic results as for Problem~\eqref{prob:2} and its special cases. In particular, we still obtain a pseudo-polynomial-time algorithm to solve Problem \eqref{prob:5} when the data is rational, and a strongly polynomial time algorithm for the special case of time-independent bounds. 
Observe that this variant allows one to enforce $s_t=\bar s_t$ at any time  $t\in\tset$ by setting $L_t^s=U_t^s=\bar s_t$.
When the lower bounds on purchases and sales are strictly positive, the complementarity constraints are equivalent to the linear constraints
\[
w_t+z_t\le 1,\qquad t\in\tset,
\]
and therefore require no additional analysis.

\if0\subsection{Multiple Vendors}

In this variant, there are multiple vendors who buy and sell the commodity and we can choose to split the amount we buy and sell among these different vendors at any time period, subject to bounds. To formalize this, consider a set $V$ of vendors. At any time $t$, let $y^i_t$ be the amount sold to vendor $i\in V$ and let $x^i_t$ be the amount purchased from vendor $i$. Let $p^i_t$ and $c^i_t$ denote the selling and purchasing prices for vendor $i$, respectively. Let $U^{y_i}_t$ be the bound on $y^i_t$ and let $U^{x_i}_t$ be the bound on $x^i_t$. We stick with the assumption that at any time $t$, all the sales precede all the purchases. Thus, the variant of Problem~\eqref{prob:1} that we obtain is the following:

\begin{align*}\label{prob:6}
	\max & \;\;\;\;\sum_{t=1}^\tbound~\left(\rt\st+\sum_{i\in V}\left(p_t^i\yt^i+c_t^i\xt^i\right)\right) \tag{\bf{LP5}}\\
	\text{s.t.}&\;\;\;\; \st = \s_{t-1} - \sum_{i\in V}\yt^i + \sum_{i\in V}\xt^i & \forall\; t\in\tset\\
	&\;\;\;\;\sum_{i\in V}\yt^i \leq \s_{t-1} &\forall\; t\in\tset \\
	&\;\;\;\; 0\leq \st\leq \Ust,\;\;\;\;0\leq \yt^i\leq U^{y_i}_t,\;\;\;\; 0\leq \xt^i\leq U^{x_i}_t& \forall\; t\in\tset,\; i\in V
\end{align*}

Let the above polytope describing the feasible region be called $P_5$. Here, the characterization of extreme points as critical points still holds, i.e., in any extreme point of $P_5$, there can be at most one vendor with $0<x^i_t<U^{x_i}_t$ or $0<y^i_t<U^{y_i}_t$ between consecutive stock-extreme times. Hence we obtain a new characterization of the set of potential stock levels at the end of time $t$ in any extreme point of $P_5$.

\begin{align}
	S''_t =&   \Big\{\big\{K-\sum_{k=1}^t \sum_{i\in V}v^i_k U^{y_i}_k + \sum_{k=1}^t\sum_{i\in V} u^i_k U^{x_i}_k\::\:K\in \{0,\snought,\Us_1,\ldots,\Ust\},\; v^i,u^i\in\{0,1\}^t\big\}\\
	&\quad\bigcup
	\big\{K' + \sum_{k=t+1}^T \sum_{i\in V}v^{i'}_k U^{y_i}_k - \sum_{k=t+1}^T\sum_{i\in V}u^{i'}_kU^{x_i}_k\::\: K'\in \{0,\Us_{t+1},\ldots,\Us_T\},\; v^{i'},u^{i'}\in\{0,1\}^{T-t}\big\}\Big\}\bigcap \Big[0,\Ust\Big].\nonumber
\end{align}

Now, suppose we introduce vendor-specific fixed costs. Let $g_t^i$ denote the fixed cost of purchasing from vendor $i$ during time period $t$ and let $h_t^i$ denote the fixed cost of selling to vendor $i$ during time period $t$. We introduce binary variables $z^i_t$ indicating whether a sale to vendor $i$ takes place during time period $t$ and $w^i_t$ indicating whether a purchase from vendor $i$ takes place during time period $t$. The resulting problem is

\begin{align*}\label{prob:7}
	\max & \;\;\;\;\sum_{t=1}^\tbound~\left(\rt\st+\sum_{i\in V}\left(p_t^i\yt^i+c_t^i\xt^i+h_t^i\zt^i+g_t^i\wt^i\right)\right) \tag{\bf{IP5}}\\
	\text{s.t.}&\;\;\;\; \st = \s_{t-1} - \sum_{i\in V}\yt^i + \sum_{i\in V}\xt^i & \forall\; t\in\tset\\
	&\;\;\;\;\sum_{i\in V}\yt^i \leq \s_{t-1} &\forall\; t\in\tset \\
	&\;\;\;\; 0\leq \st\leq \Ust,\;\;\;\;0\leq \yt^i\leq U^{y_i}_t\zt^i,\;\;\;\; 0\leq \xt^i\leq U^{x_i}_t\wt^i& \forall\; t\in\tset,\; i\in V\\
	&\;\;\;\; \zt^i,\wt^i\in\{0,1\} & \forall\; t\in\tset,\; i\in V
\end{align*}

As in Lemma~\ref{lem:fixedcosts}, we can show that the extreme points of the feasible region are extensions of the extreme points of $P_5$. Hence, using the above description, we know a set of potential stock levels in any extreme point allowing us to create the network flow formulation. The $(x,y,z,w)$ values on any edge connecting $\Bar{s}_{t-1}$ and $\Bar{s}_t$ will be described by the solution to the following integer program:

\begin{align*}
	\max & \;\;\;\;\sum_{i\in V}\left(p_t^i\yt^i+c_t^i\xt^i+h_t^i\zt^i+g_t^i\wt^i\right)\\
	\text{s.t.}&\;\;\;\; \Bar{s}_t = \Bar{s}_{t-1} - \sum_{i\in V}\yt^i + \sum_{i\in V}\xt^i\\
	&\;\;\;\;\sum_{i\in V}\yt^i \leq \Bar{s}_{t-1}\\
	&\;\;\;\;0\leq \yt^i\leq U^{y_i}_t\zt^i,\;\;\;\; 0\leq \xt^i\leq U^{x_i}_t\wt^i& i\in V\\
	& \;\;\;\; \zt^i,\wt^i \in \{0,1\} & i\in V.
\end{align*}

Note that if the number of vendors is fixed, then the above integer program has a constant number of variables and constraints. Consequently, it can be solved in constant time.
\fi
\subsection{{Separable Convex Objectives}  }

Finally, we observe that the linear objective functions considered throughout the paper can be replaced by separable convex pay-off functions.
Suppose,  in \eqref{prob:4} the pay-off at time $t$ is given by a convex function of the variables
$s_t,x_t,y_t,w_t,z_t,$ and $v_t$, rather than a linear function.
We then modify problem Problem \eqref{prob:4} by changing the objective function to
\[\max \sum_{t=1}^T u_t(s_t,x_t,y_t,w_t,z_t,v_t)\]
Here $u_t$ are known convex functions of six variables. In this setting, the objective function is a sum of convex functions and therefore it is convex. Since we are maximizing a convex function over a bounded polyhedron, there must exist an extreme-point optimal solution.

From the earlier sections, we already know that the extreme points of the feasible region $P_3$ are given by extensions of critical points of $\Pone$. Also, critical points of $\Pone$ correspond to paths of the network $G_1$. 
Hence, it suffices to modify the arc weights in $G_1$ to reflect the convex pay-off functions so that a longest path in the network $G_1$ would correspond to an optimal solution to our problem. This is indeed easy to do since on any arc of $G_1$ associated with a time period $t$, the values of $x_t, y_t$ and $s_t$ are fixed. We can therefore select $w_t,z_t$ and $v_t$ in such a way that $u_t(s_t,x_t,y_t,w_t,z_t,v_t)$ is maximized while maintaining feasibility. Since there are at most eight possible values for the triple $(w_t,z_t,v_t)$, the corresponding arc weight can be computed in constant time. Note that one can also impose the complementarity constraints $x_ty_t = 0$ for all $t\in \tset$ since the extreme points of the resulting feasible region are characterized by the network $G_2$.
Thus, Algorithms~1 and~2 extend naturally to the case of separable convex pay-off functions. Their running times remain unchanged, both in the general case and in the special cases discussed in Section~\ref{sec:special}.

\section{Conclusions and Future Work}\label{sec:conclude}

The warehouse problem with fixed costs, complementarity constraints, or both is a difficult problem to solve in its most general setting. However, most of the difficulty arises due to the presence of time-varying bounds. In most applications of the warehouse problem, the bounds are assumed to be time-independent and the results in this paper provide strongly polynomial algorithms to solve the warehouse problem in those settings. Additionally, the exact characterization of extreme points of polytopes $\Pone,\Ptwo$ and $\Pthree$ as critical points is very useful as has already been observed in Section \ref{sec:further}. 
The characterization of extreme points developed in this paper provides a common framework for constructing network-flow formulations and extended formulations for a broad class of warehouse models. There are also some promising future directions that one can explore, some of which are:

\begin{enumerate}
	\item The general problems \eqref{prob:2} and \eqref{prob:3} are \textit{NP}-hard and it would be helpful to develop approximation algorithms to solve these problems.
	
	\item This paper characterizes the extreme points of the polytopes $\Ptwo$ and $\Pthree$. It would be helpful to also obtain a characterization or description of the facet-defining inequalities for these polytopes.
	
	\item This paper discusses the optimization problem for polytopes $\Ptwo$ and $\Pthree$ and under certain reasonable settings provides polynomial time algorithms with decent running times. It would be helpful to also have fast separation algorithms for these polytopes in those settings.
	
	\item Consider the multi-commodity warehouse problem where the owner of the warehouse trades multiple commodities all sharing the same warehouse. One can attempt to extend the results of this paper in those settings as well.
\end{enumerate}

\bibliography{ref} \bibliographystyle{plain}

\end{document}